# Factors Influencing the Quality of the User Experience in Ubiquitous Recommender Systems

Nikolaos Polatidis, Christos K. Georgiadis

Department of Applied Informatics, University of Macedonia, Thessaloniki, Greece
{polatidisn, gxri}@acm.org

**Abstract.** The use of mobile devices and the rapid growth of the internet and networking infrastructure has brought the necessity of using Ubiquitous recommender systems. However in mobile devices there are different factors that need to be considered in order to get more useful recommendations and increase the quality of the user experience. This paper gives an overview of the factors related to the quality and proposes a new hybrid recommendation model.

## 1 Introduction

Recommender systems are software algorithms aiming at filtering information [7]. Their job is to propose items or services using information based on user preferences. Recommender systems main algorithms are based on collaborative filtering, which is the most widely used algorithm. The items or services are recommender according to preferences of other users that have similar preferences [7]. Another important recommendation algorithm is content based filtering where the recommendations depend on previous items found in the history of the user and the top matching are proposed by the system [7]. Further recommendation algorithms include knowledge based filtering where the system uses a knowledge based attitude to generate recommendations. It is an algorithm where the user pre defines a set of requirements that the system will use to create the list of the recommendations. Moreover the knowledge database can be built by recording the user preferences while he is browsing or by asking him to complete a questionnaire [7].

Hybrid recommender systems use a combination of the above methods and look the most promising due to the fact that can take the advantage of each method and improve the overall output. The hybridization can occur in different ways such as using the output of one algorithm as the input for the other or by combining the recommendations of each algorithm at the interface level [7].

Ubiquitous recommender systems assist the user of a mobile device by providing him with personalized recommendations of items or services that are in the proximity

[9]. These recommendations usually include mobile tourism related services such as tourist guides, shopping recommenders and route finders [9], [13]. A clear example of ubiquitous recommendations can be found in [15] where a city guide is proposed by the authors for mobile device users that are equipped with GPS in their devices. Moreover it has been proposed that ubiquitous recommender systems can make smoother the buying process in the actual store by recommending items that are of the user interest [12]. Such recommenders can suggest items, display their ratings and comments.

The idea of ubiquitous computing as proposed by Want and Pering [16] and the main idea was to move away from traditional desktop environments to distributed computing, using a variety of devices. In addition it usually referred as pervasive computing [19], [16]. A critical part of ubiquitous recommendations is context awareness that has to be taken into consideration in order to provide accurate recommendations [2], [3], [9]. This brings us to a critical point where if we want to have quality recommendation we have to let the system use the location and at the same time have our privacy respected. Such systems aim to solve the information overload problem found nowadays on the internet and do it successfully up to a point. However different quality factors have to be ensured in order to improve the user experience and increase the overall quality.

The diagram below gives an overview of a ubiquitous context aware recommender.

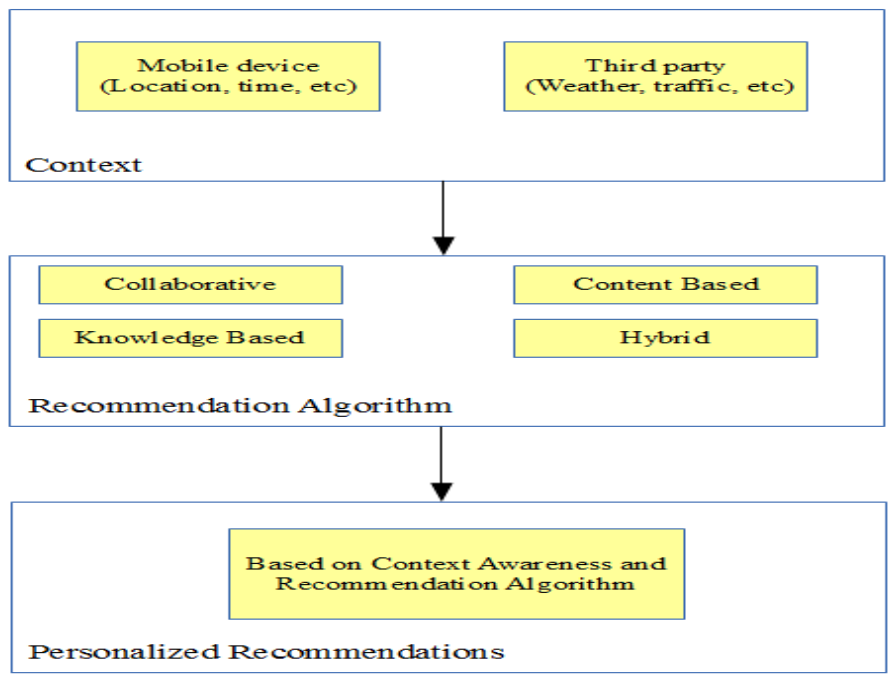

**Fig. 1.** Ubiquitous context aware recommender system

This work is focused on introducing ubiquitous recommender systems and how they have emerged and then a detailed description of the factors that affect the user experience follows. Although a number of different technologies has to be used the key challenge is to hide the presence of such technologies from the user and develop a smooth integrated process, which will include all the related factors as an integrated and complete framework that aims to deliver the right content to the user appropriately and securely. Moreover, the overall aim of the research is to highlight the factors and suggest that it would be beneficial to the community if different people from diverse contexts could benefit, while being in different environments, such as office, home and public spaces.

## 2   Influencing Factors

Main factors include context awareness, privacy and algorithms [9]. Also a challenge that is found in traditional recommender systems but also applies to ubiquitous recommendations is the 'new user' problem, which is an important factors that plays a vital role in the development of such systems. It is noted that the factors that affect considerably the quality of the user experience in ubiquitous recommender systems are not found in other environments and are primarily to the size of the device, the physical resources and the amount of time the user is willing to use a small size device. Furthermore a less critical factor but considered essential is multilingual personalization [6].

### 2.1   Context Awareness

Context can be used by ubiquitous recommender systems to produce more personalized recommendations [1]. Recommender systems use collaborative and content filtering methods most of the time to produce recommendations, however this methodology does not take into consideration the contextual information and how this can be applied to the current situation and increase the overall quality of recommendations. According to the same scholars contextual recommender systems can be categorized in three main types. Fully observable, partially observable and unobservable. Moreover, a point is to discover the changes in the contextual factors and how to represent them in a mobile environment.

Ubiquitous context aware recommender systems vary and include different factors such as location, time, weather and emotional status of the user. The contextual information is very important if we want to provide recommendations that are based on Location Based Services [1].

Contextual information can be collected either explicitly, which is by asking the user directly to provide data using a questionnaire. Moreover data can collected implicitly by environment data, such as historical information and changes that occur during the use of the service [1]. Required values may be taken into the system by using the sensors of the device such as the camera and the Global Positioning System (GPS) [5].

Context is considered to be the most important aspect in ubiquitous recommender systems [1], [9]. We strongly believe that if context is utilized properly more useful recommendations will occur and the user will be highly satisfied.

### 2.2 Privacy

Privacy means that the user is ensured and decides on what ways his data will be processed [8], [10]. Privacy concerns direct users towards a negative behavior when they are asked to provide more data in order to receive personalized recommendations.

In Recommender Systems users are divided in three main categories [10]:

- Users that will provide any kind of information in exchange with the highest level of personalization possible.

- Users that will give some information so they can receive some kind of personalized recommendations.

- Users that will not give any kind of information due to privacy concerns.

Privacy is crucial factor that it is possible to be addressed using the right techniques. If this issue didn't exist then the user would supply any necessary information and his experience using the recommender system would be of a very high standard.

### 2.3 New user and item

The new user and item problem are very important when the algorithm used is based solely in collaborative filtering (CF). They occur when a new user or item is added to the database there is no history about the user or no rating history about the item or service.

If a user wants the highest quality possible from a recommender then this is a very important issue that needs to be faced and this can only be dealt with the use of hybridization techniques.

Hybrid recommender are divided into three main categories [7]:

- Parallel

- Monolithic

- Pipelined

Parallel hybrid recommender use the same input in one or more recommendation algorithms and then combine the output of each into a hybridization step and produce a single output. Monolithic recommenders use a single recommender and combine different techniques. Finally pipelined recommender use the output from a single recommendation algorithm as the input for the next one.

Hybrid recommender can increase the quality of recommenders overall, since single algorithms have limitations, such as the new user and item problem described above.

## 3    Less Influencing Factors

Less important factors or else defined as challenges can be found in the literature as well.

### 3.1    Perceived accuracy

A factor that needs some consideration is perceived accuracy which is a point where a user feels that the recommendations match his preferences [11]. It is considered to be a measuring assessment of how good the recommender performed and how accurate is to find the interests of a particular user.

### 3.2    Familiarity and novelty

Familiarity is a description of the previous experience that user has with the recommended item or services [11]. However familiarity might mean that all the recommendation categories must be familiar to the user. Novelty must be introduced and balanced with familiarity so the user would be as satisfied as possible.

### 3.3    Attractiveness

Attractiveness is conserved with the process of irritating the user and evoke positive imaginations and increase the possibility of desiring. Attractiveness is concerned on how well the recommendations will be delivered to the user and not the recommendations itself [11].

### 3.4    User interfaces

Limitations found in the user interface, where different devices may be used, the task would be to develop suitable and user friendly interfaces [5]. User interfaces are tightly related to the attractiveness as described above and could improve the quality. The more attractive is the user interface the user will be satisfied.

### 3.5 Multilingual Personalization

Given the fact that there is a vast amount of data found on the internet, these data can exist in different languages [6]. It is possible that the data requested from a user will not be available in his native language but be available in a foreign language.

Research has been done towards the field of personalized information retrieval [6]. It is a field where if suitable research occurs then more useful recommendations will be delivered.

## 4 Proposed Model

User experience is more and more becoming an essential part in the attention of the research community. However there isn't much work done on how the quality of the user experience in ubiquitous recommender systems can be increased and what kind of standards could be specified to work towards that direction.

The criteria need to be combined into a comprehensive framework that could be potentially used to evaluate the qualities of ubiquitous recommender systems. The framework should take into consideration all the major criteria that should be satisfied. A comprehensive model identifying all the aforementioned essential qualities must be established as a standard. That is how potential users will adapt a system.

Moreover it should be noted that the quality of recommendations and hence an increased user experience is heavily based on the algorithm used. A hybrid algorithm based on collaborative filtering is necessary due to the better prediction of hybrid algorithms. However there is a gap in collaborative filtering with the new user and item issues, which can be solved with the use of data from social media such as Facebook. In addition the proposed algorithm must incorporate contextual information to be useful in ubiquitous environments. The figure below gives an abstract representation of the model.

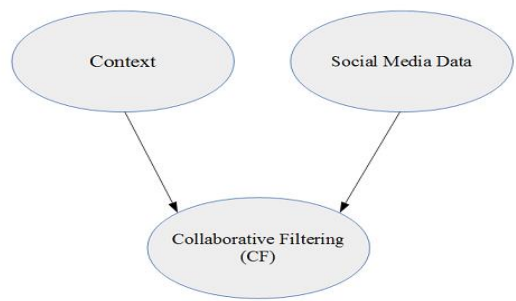

**Fig. 2.** Proposed hybrid model abstract representation

An issue is that in a social network different types of relationships can be found. Consider the graph and the table below that are two different social relationships.

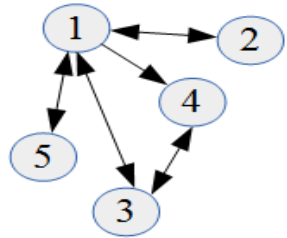

**Fig. 3.** Graph representation of a social relationship

In the above network there are different kind of relationships such as friendship, which is denoted by a double direction arrow and follower or member, which is denoted by single direction arrow.

At the table UxU below where a zero value is denoted it means that there is no kind of relationship and where one is denoted there is some kind of relationship.

| User1 | User2 | User3 | User4 | User5 | User6 |
|---|---|---|---|---|---|
| User2 | 0 | 1 | 1 | 1 | 0 |
| User3 | 1 | 0 | 0 | 0 | 1 |
| User4 | 1 | 0 | 0 | 1 | 1 |
| User5 | 0 | 1 | 0 | 0 | 0 |
| User6 | 1 | 1 | 0 | 0 | 0 |

**Table 1.** Representation of socials relationships

The table below describes a ratings database of users about items or services that could be found in a social network.

|  | Item1 | Item2 |
|---|---|---|
| User1 | 2 | 5 |
| User2 | 0 | 6 |
| User3 | 5 | 5 |
| User4 | 2 | 1 |

**Table 2.** User ratings of items in a social network

However the larger the network gets then it will become very difficult to identify relationships between that impose an actual value so they can be used. We chose the *K-means* algorithm [14] as the clustering algorithm due to its simplicity. K is the desirable number of clusters we want.

The users then will be clustered into N groups so the algorithm will be searching only on relevant user data.

The set of users is represented as follows:

U = {User1, User2, User3, .... , User n}

The clusters are then represented as follows:

C = {Cluster1, Cluster2, Cluster3, ... Cluster n}

Each cluster C is a set of user that are related.

Cluster1 = {u1, u2, u3, ... , un}

The proposed algorithm is described below and utilizes the k-means clustering.

---
**Algorithm** Social-Media-Clustering
**Input**
U: the set of people and groups related to the user
G: the graph of the user that defines the relationships
TR: the matrix with the user ratings
**Output**
C: The superset of the required cluster sets

---

**Fig. 4.** Social media clustering algorithm

The k-means algorithm outline is as follows:

---
**N** is the number of total users
**K** specifies the number of the required clusters
**Sim** (x, y) is the similarity function that will be used by k-means
Take the first k users and assign them as centroids
Compare the rest users to the centroid users
Assign users to clusters

---

**Fig. 5.** Outline of the K-means algorithm

We will use the Pearson correlation as the similarity measure function. The similarity between users i and j is defined as follows:

$$sim(i,j) = \frac{\sum_{u \in U}(R_{u,i} - \bar{R}_i)(R_{u,j} - \bar{R}_j)}{\sqrt{\sum_{u \in U}(R_{u,i} - \bar{R}_i)^2}\sqrt{\sum_{u \in U}(R_{u,j} - \bar{R}_j)^2}}$$

**Fig. 6.** Pearson correlation

## 5      Conclusions and Future Work

Recommender systems has matured to a full research area both in academia and in practice. However extended research has still to be done in ubiquitous environments and as the field grows, significant, new, challenges will be faced in terms of infrastructure and criteria. This is due to the fact that two different areas have to be researched and as ubiquitous computing and recommender system develop further many more characteristics will appear and new solutions will have to be proposed. Ubiquitous recommender systems will have to combine different characteristics to become useful to our everyday lives and provide an improved user experience.

Quality is a very important aspect found everywhere, including recommenders and ubiquitous environments. It is vital for the designer to be aware of the factors that relate to the improvement of the user experience. Most important factors that need to be addressed include context awareness, privacy and the new user and item problem found in collaborative filtering. Less important factors that if addressed could potentially improve the quality of the user experience include perceived accuracy, familiarity and novelty, attractiveness, improved user interfaces and multilingual personalization.

However mobile devices and networking infrastructures are evolving constantly and new challenges arise. Both designers and developers should be aware of new open problems and implications.

In addition it should be noted that although it is an important research field there is not much work in the literature regarding quality and serious work should be taken to define the required criteria that need to be satisfied.

Also simple case studies can be employed with the form of a questionnaire so that results from real users can be used to design better systems. The case studies should use the precision and recall methods [4]. These studies could be employed both in the prediction of the algorithm in retrieving the relevant friends and for the actual prediction system.